\pdfoutput=1
\documentclass[11pt]{article}
\usepackage{graphicx}
\usepackage{bm}
\usepackage{color}
\catcode`@=12
\newcommand{\be}{\begin{equation}}
\newcommand{\ee}{\end{equation}}
\newcommand{\bea}{\begin{eqnarray}}
\newcommand{\eea}{\end{eqnarray}}
\def\r{\rightarrow}
 \def\b{\lambda}

\def\s12{\sin\theta_{12}}
\def\s23{\sin\theta_{23}}
\def\s13{\sin\theta_{13}}
\def\ts12{\theta_{12}}
\def\ta23{\theta_{23}}
\def\t13{\theta_{13}}

\makeatletter
\@addtoreset{equation}{section}

\makeatother
\title{\LARGE\bf Maximal zero textures of the inverse 
seesaw with broken $\mu\tau$ symmetry
\footnote
{To be published in Indian Journal of Physics (Special issue 
commemorating the 125th birth anniversary of C.~V.~Raman).}}
\author{{\bf Biswajit Adhikary$^{\rm a}$\footnote{biswajit.adhikary@saha.ac.in}
     , Ambar Ghosal$^{\rm b}$\footnote{ambar.ghosal@saha.ac.in} and Probir Roy$^{\rm b}$
\footnote{probir.roy@saha.ac.in}}\\
  a)Department of Physics, Gurudas College,
Narkeldanga, Kolkata-700054, India\\
  b) Saha Institute of Nuclear Physics, 1/AF Bidhannagar, 
 Kolkata 700064, India}
\date{}
\begin{document}
\maketitle
\thispagestyle{empty}
\begin{abstract}
\noindent
 The inverse neutrino seesaw, characterised by only one source of lepton
 number violation at an ultralight $O$(keV) scale and observable new
 phenomena at TeV energies accessible to the LHC, is considered. Maximal
 zero textures of the $3\times 3$ lighter and heavier Dirac mass matrices of
 neutral leptons, appearing in the Lagarangian for such an inverse seesaw, are
 studied within the framework of $\mu\tau$ symmetry in a specified weak
 basis. That symmetry ensures the identity of the positions of maximal zeros
  of the heavy neutrino mass matrix and its inverse.
 It then suffices to study the maximal zeros of the lighter Dirac mass 
 matrix and
 those of the inverse of the heavier one since they come
 in a product. The observed absence of any unmixed neutrino flavour and the
 assumption of no strictly massless physical neutrino state 
allow only eight $4$-zero
 $\times$ $4$-zero,  eight $4$-zero $\times$ $6$-zero and eight $6$-zero
 $\times$ $4$-zero combinations. The additional requirement of leptogenesis is
 shown to eliminate the last sixteen textures. The surviving eight $4$-zero
 $\times$ $4$-zero textures are subjected to the most general explicit
 $\mu\tau$ symmetry breaking terms in the  Lagarangian in order 
to accommodate 
the nonzero value of $\theta_{13}$ in the observed
 range. 
A full diagonalisation is then carried out.
On numerical comparison with all extant and 
relevant neutrino (antineutrino)
 data, seven of these eight combination textures in five neutrino matrix 
forms are found to be allowed, leading to five distinct neutrino mass 
matrices.
 Two of these permit only a normal (and the other three only an inverted) 
mass ordering of the light neutrinos.
\end{abstract}
\vskip 0.1in
%%%%%%%%%%%%%%%%%%%%%%%%%%%%%%%%%%%%%%%%%%%%%%%%%%%%%%%%%%%%%
\section{Introduction}
Type-I seesaw \cite{Gell,Yana,Mo} has been the most popularly used mechanism so far to generate
light neutrino masses. Unfortunately, all new phenomena in this scenario
involve three heavy right chiral $SU(2)_L\times U(1)_Y$ 
singlet neutrinos $\nu_{kR}$
($k=e,~\mu,~\tau$ and $R$ a chiral index) with a Majorana 
mass matrix $M$, the smallest
eigenvalue of which is bounded \cite{four} from below by 
the leptogenesis constraint to 
$M_{lightest}\ge 10^8$ GeV. Thus they are beyond
the foreseeable reach of laboratory experiments. Alternative ways exist of
circumventing  this. In particular, there is the inverse 
seesaw mechanism 
\cite{Mohapatra:1986aw,Mohapatra:1986bd,Bernabeu:1987gr} 
to
which we direct attention. 
Three additional left chiral $SU(2)_L\times
U(1)_Y$ singlet fields $s_{L}$ are introduced here along with their 
$3\times 3$ Majorana mass matrix
$\mu$ characterised by only ultralight eigenvalues in the keV scale. This 
mass matrix constitutes the sole
source of lepton number violation since three other $3\times 3$ lepton mass
matrices, which are introduced, are of a Dirac type: $M_\ell$ between ${\bar
l}_L$ and $l_R$ ($l$ being a charged lepton), $M_D$ between ${\bar
\nu}_L$ and $\nu_R$ and $M$ between ${\bar s}_L$ and $\nu_R$. Here $M_D$ can
be chosen with nonzero elements of the order of the 
 known charged fermion masses, 
while those
of $M$ can be taken to be at the TeV scale with phenomenological consequences
of interest to the LHC. Thus we have $O(\mu)\ll O(m_D)\ll O(M)$. 
Leptogenesis has been shown to be \cite{six} realistically
possible in this scenario. 
\par
The key feature of this inverse seesaw, as shown below, is that the product 
\begin{equation}
F \equiv M_D M^{-1}
\label{one}
\end{equation}
plays a role in the light neutrino mass formula which is analogous to 
that of the Dirac mass matrix $M_D$ in type-I seesaw.
We also consider the question of texture. 
A fruitful approach to the problem of light neutrino masses
and mixing angles has been based on the idea \cite{seven} of maximal texture
zeros. In the present case these vanishing elements are postulated 
in the matrices $M_D$ and $M^{-1}$. Texture statements being 
basis dependent, one needs to choose a weak
basis \cite{brancobook} 
in order to make such a statement. 
We choose one in which the flavoured charged leptons $l$ and the
singlet fields $s_L$ are mass diagonal with real and positive entries. 
Such a
choice can be consistently made since these two sets of fields do not have any
mutual interaction at the Lagrangian level. 
In such a weak basis, in analogy with the type-I seesaw case 
\cite{seven}, four and six turn out to be the number of maximal zeros 
allowed respectively in $M_D$ and $M^{-1}$ which are the two factors in the 
matrix $F$. In order to simplify the possible textures, we assume $\mu\tau$ 
symmetry (with a small breaking), motivated by the near-maximal and small 
values of $\theta_{23}$ and $\theta_{13}$ respectively. The further assumption 
of no strictly massless neutrino, 
bolstered by the observed fact of a nontrivial 
mixing for every neutrino flavor, allows only eight 4-zero $\times$ 6-zero, 
eight 6-zero $\times$ 4-zero and eight 4-zero $\times$ 4-zero combinations. 
Of these, the first sixteen cannot effect leptogenesis and hence are ruled out. Of the remaining eight, 
seven combination textures, leading to five distinct neutrino matrix 
forms, are found to be allowed.
 Two of these allow only a normal (and the other three only an inverted) 
mass ordering of the light neutrinos after comparison with all available 
neutrino and antineutrino data. 
\par
In the rest of the paper, 
Sec. 2 is devoted to a discussion of maximal zero textures with 
$\mu\tau$ symmetry within the framework of the inverse seesaw. Section 3
 shows how the requirement of leptogenesis constrains these textures. The 
effect of $\mu\tau$ symmetry  breaking on the allowed texture combinations
 is discussed 
in Sec.4. The diagonalisation of the resultant neutrino mass matrices 
with broken $\mu\tau$ symmetry is carried out in Sec. 5. 
The numerical analysis in comparing the predictions of these surviving 
mass matrices and a global set of neutrino and antineutrino data is briefly 
presented in Sec.6. In Sec. 7 we summarise our conclusions. In the 
Appendix some relevant expressions, used in the text, are given explicitly.  

\section{Inverse seesaw, maximal zeros and $\mu\tau$ symmetry}
The Lagrangian mass terms, required to facilitate the invesre seesaw, can be
written as follows:
\begin{eqnarray}
-{\cal L}^{\rm mass}&=&\overline{l_{kL}}(M)_{ll'}l_{k'R}
 +\overline{\nu_{kL}}(M_D)_{kk'}\nu_{k'R}+\overline{s_{kL}}M_{kk'}\nu_{k'R}\nonumber\\
&&+\frac{1}{2}\overline{s_{kL}}\mu_{kk'}s^C_{k'R}+h.c.\nonumber\\
&=&\overline{l_{kL}}(M)_{ll'}l_{k'R} + \frac{1}{2}
\left(\begin{array}{ccc}{\overline{\nu_L}}&{\overline{{(\nu^C)}_L}}&\overline{
 s_L} \end{array}\right)\left(\begin{array}{ccc} 0 & M_D & 0\\ M_D^T & 0 &
 M^T\\ 0 & M
 & \mu \end{array} \right)\left(\begin{array}{c}{(\nu^C)}_R \\ \nu_R \\
 {(s^C)}_R\end{array}\right)  + h.c\nonumber \\
\label{invlag}
\end{eqnarray}  
The magnitudes of the elements of the three mass matrices introduced 
above obey a hierarchy
$O(\mu)\ll O(M_D)\ll O(M)$, as mentioned in the Introduction. The inverse
seesaw is then effected by the $3\times 3$ block matrix in the second RHS term 
of
 (\ref{invlag}), leading to the light neutrino mass terms 
\begin{eqnarray}
-{\cal L}^{\nu-{\rm mass}}=\overline{\nu_{L}}M_\nu{(\nu^C)}_R + h.c.
\label{nmt}
\end{eqnarray} 
In (\ref{nmt}), the effective light neutrino mass 
matrix $M_\nu$ is given by the  inverse seesaw formula 
\begin{eqnarray}
M_\nu\simeq M_DM^{-1}\mu(M^T)^{-1}M_D^T\equiv F\mu F^T.
\label{invmas}
\end{eqnarray}
The interplay among the widely different energy scales involved 
in (\ref{invmas}) is
manifest in the relation
\begin{eqnarray}
\left(\frac{M_\nu}{0.1{\rm eV}}\right)\sim{\left(\frac{M_D}{10^2{\rm {GeV}}}
\right)}^2\left(\frac{\mu}{1{\rm{keV}}}\right)
{\left(\frac{M}{10^4{\rm{GeV}}}\right)}^{-2},
\label{F}
\end{eqnarray}
where $M_\nu$, $M_D$, $\mu$ and $M$ are typical elements of the 
corresponding matrices.
In our chosen weak basis, we have 
 \begin{eqnarray}
M_\ell={\rm diag}\left(m_e,~m_\mu,~m_\tau\right),~~\mu=
{\rm diag}\left(\mu_1,~\mu_2,~\mu_3\right).
\label{dmlmu}
\end{eqnarray}
\par
Two additional inputs are now introduced to  restrict the form of
$M_\nu$. First, there is the observed fact that none of the three known 
light neutrinos is free from flavour mixing. 
This means that any block diagonal form
of $M_\nu$ is inadmissible. The same statement can be made about 
the matrix $F$ of (\ref{one}) on account of (\ref{invmas}). 
Indeed, if any row in a texture of $F$ 
  is orthogonal to both the other rows, one neutrino family
  decouples -- disallowing that texture. Second, we assume that no 
neutrino is
  strictly massless because of the lack of any fundamental principle 
(unlike for a photon)
 suggesting that such be the case. Since that leads to  
a nonvanishing ${\rm
 det}\, M_\nu$, (\ref{invmas}) implies that both  ${\rm det}\, M_D$ and  
${\rm det}\, \mu$ have to be nonvanishing. 
Thus not only is each $\mu_i$  nonzero for $i=1,2,3$,
 neither $M_\nu$ nor $M_D$ can have a vanishing row/column or a quartret of
 zeros at the four corners of a rectangular array. 
The same statement can be made
 about the matrix $F$ of (\ref{F}). The analysis, detailed in 
Ref. \cite{Adhikary:2009kz}-\cite{Adhikary:2012zx} 
for 
$M_D$, now
 holds {\it mutatis mutandis} for $F$. Therefore, the maximum number of zeros
 that $F$ can accommodate is four. Moreover, there are seventy two such allowed
 textures of $F$ with four vanishing entries. 
\par
We propose to reduce the number of these seventy-two four zero 
textures of $F$ with the aid of $\mu\tau$ symmetry. This
symmetry{\footnote{For a review and original references, 
see Ref. \cite{mutaureview}}}
postulated only in the neutrino sector, stipulates an invariance of all
neutrino terms in the Lagarangian under the interchange of the flavor indices
$2\leftrightarrow 3$. This automatically requires $\mu_2=\mu_3$ and moreover
implies that 
\begin{eqnarray}
\Phi_{22}=\Phi_{33},~\Phi_{12}=\Phi_{13},~\Phi_{21}=
\Phi_{31},~\Phi_{32}=\Phi_{23},
\label{mtmas}
\end{eqnarray}
where $\Phi$ is any of the matrices $M_D$, $M$, $M^{-1}$, $F$. We then 
deconstruct the surviving textures in terms of those of $M_D$ and $M^{-1}$. 
The 
$\mu\tau$ symmetric forms of $M_D$ and $M$ can be written as
\begin{eqnarray}
M_{D}=\left(\begin{array}{ccc} A_1 & A_2 & A_2\\ B_1 & B_2 & B_3\\B_1 & B_3 & B_2 \end{array}
\right),\qquad
M=\left(\begin{array}{ccc} r_1 & r_2 & r_2\\ s_1 & s_2 & s_3\\s_1 & s_3 & s_2 
\end{array}
\right),
\label{mtmdm}
\end{eqnarray}
where the entries $A_{1,2}$, $B_{1,2,3}$, $r_{1,2}$ and $s_{1,2,3}$ are 
unknown dimensional complex quantities in general.
 
It was shown  \cite{seven} 
in the context of type-I seesaw that the number of allowed
$4$-zero Yukawa textures gets drastically reduced from seventy two to four on
the imposition of $\mu\tau$ symmetry. In case of the inverse seesaw, an
identical statement holds for the matrix $F$. The four textures, divided as in
the earlier case into two categories $A$ and $B$, are
\begin{eqnarray}
F^{A1}=\left(\begin{array}{ccc} a_1 & a_2 & a_2\\ 0 & b_2 & 0\\0 & 0 & b_2 \end{array}
\right),\qquad
F^{A2}=\left(\begin{array}{ccc} a_1 & a_2 & a_2\\ 0 & 0 & b_2\\0 & b_2 & 0 \end{array}
\right),\nonumber\\
 F^{B1}=\left(\begin{array}{ccc} a_1 & 0 & 0\\ b_1 & b_2 & 0\\b_1 & 0 & b_2  \end{array}
\right),\qquad
F^{B2}=\left(\begin{array}{ccc} a_1 & 0 & 0\\ b_1 & 0 & b_2\\b_1 & b_2 & 0 \end{array},
\right),
\label{al40}
\end{eqnarray}
where $a_{1,2}$ and $b_{1,2}$ can be given in terms of 
$A_{1,2}$, $B_{1,2}$, $r_{1,2}$. Once again, 
the two textures of Category $A$ yield an identical $M_{\nu}$ and the
same goes for the two textures of Category $B$.

An interesting consequence of $\mu\tau$ symmetry is that the positions 
of the
maximal zeros of any texture of $M$ and $M^{-1}$ become identical. 
We utilise this in the
deconstruction of each allowed $4$-zero texture of $F$ in terms of the
maximally allowed zero textures of $M_D$ and $M$. A straightforward but
tedious amount of algebra leads to the deconstruction 
listed in Table 1. Two allowed $4$-zero $\times$ $4$-zero textures, 
two allowed $4$-zero $\times$ $6$-zero textures and two 
allowed $6$-zero $\times$ $4$-zero textures
follow for {\it each} of $F^{A1}$, $F^{A2}$, $F^{B1}$ and  $F^{B2}$. 
%**********************************************************************
%\begin{longtable*}
\begin{table}[!htb]
\caption{\label{tabxtot} Deconstructed maximal zero textures of $F^{A1}$ and
$F^{A2}$ in category $A$.} 
\begin{tabular}{|c|c|c|c|c|}
\hline
$F$ & Type & $M_D$ & $M$ & Relations\\
\hline
&$4-$zero$\times$$4-$zero&$\left(\begin{array}{ccc} A_1 & A_2 & A_2\\ 0 & B_2 & 0\\0 & 0 &
B_2 \end{array}\right)$&
$\left(\begin{array}{ccc} r_1 & r_2 & r_2\\ 0 & s_2 & 0\\0 & 0 &
s_2 \end{array}\right)$
&
 $\begin{array}{c} a_1=A_1r_1^{-1} \\
 a_2=(A_2r_1-A_1r_2)r_1^{-1}s_2^{-1}\\b_2=B_2s_2^{-1}\end{array}$\\
%&&&\\
\cline{3-5}
$F^{A1}~=~$&&$\left(\begin{array}{ccc} A_1 & A_2 & A_2\\ 0 & 0 & B_3\\0 & B_3 &
0 \end{array}\right)$&
$\left(\begin{array}{ccc} r_1 & r_2 & r_2\\ 0 & 0 & s_3\\0 & s_3 &
0 \end{array}\right)$
&
 $\begin{array}{c} a_1=A_1r_1^{-1} \\
 a_2=(A_2r_1-A_1r_2)r_1^{-1}s_3^{-1}\\b_2=B_3s_3^{-1}\end{array}$\\
%&&&&\\
\cline{2-5}
$\left(\begin{array}{ccc} a_1 & a_2 & a_2\\ 0 & b_2 & 0\\0 & 0 & b_2 \end{array}
\right)$&$4-$zero$\times$$6-$zero&$\left(\begin{array}{ccc} A_1 & A_2 & A_2\\ 0 & 0 & B_3\\0 & B_3 &
0 \end{array}\right)$&
$\left(\begin{array}{ccc} r_1 & 0 & 0\\ 0 & 0 & s_3\\0 & s_3 &
0 \end{array}\right)$
&
 $\begin{array}{c} a_1=A_1r_1^{-1} \\
 a_2=A_2s_3^{-1}\\b_2=B_3s_3^{-1}\end{array}$\\
\cline{3-5}
&&$\left(\begin{array}{ccc} A_1 & A_2 & A_2\\ 0 & B_2 & 0\\0 & 0 &
B_2 \end{array}\right)$&
$\left(\begin{array}{ccc} r_1 & 0 & 0\\ 0 & s_2 & 0\\0 & 0 &
s_2 \end{array}\right)$
&
 $\begin{array}{c} a_1=A_1r_1^{-1} \\
 a_2=A_2s_2^{-1}\\b_2=B_2s_2^{-1}\end{array}$\\
\cline{2-5}
&$6-$zero$\times$$4-$zero&$\left(\begin{array}{ccc} A_1 & 0 & 0\\ 0 & B_2 & 0\\0 & 0 &
B_2 \end{array}\right)$&
$\left(\begin{array}{ccc} r_1 & r_2 & r_2\\ 0 & s_2 & 0\\0 & 0 &
s_2 \end{array}\right)$
&
 $\begin{array}{c} a_1=A_1r_1^{-1} \\
 a_2=-A_1r_2r_1^{-1}s_2^{-1}\\b_2=B_2s_2^{-1}\end{array}$\\
%&&&&\\
\cline{3-5}
&&$\left(\begin{array}{ccc} A_1 & 0 & 0\\ 0 & 0 & B_3\\0 & B_3 &
0 \end{array}\right)$&
$\left(\begin{array}{ccc} r_1 & r_2 & r_2\\ 0 & 0 & s_3\\0 & s_3 &
0 \end{array}\right)$
&
 $\begin{array}{c} a_1=A_1r_1^{-1} \\
 a_2=-A_1r_2r_1^{-1}s_3^{-1}\\b_2=B_3s_3^{-1}\end{array}$\\
\hline
%\end{tabular}
%\end{ruledtabular}
%\end{table}

%\begin{longtable*}
%\begin{table}[!ht]
%\caption{\label{tabxtot}  Deconstructed maximal zero textures for $F^{A2}$.} 
%\begin{tabular}{|c|c|c|c|c|}
%\hline
%$F^{A2}$ & Type & $m_D$ & $M$ & Relations\\
\hline
&$4-$zero$\times$$4-$zero&$\left(\begin{array}{ccc} A_1 & A_2 & A_2\\ 0 & 0 & B_3\\0 & B_3 &
0 \end{array}\right)$&
$\left(\begin{array}{ccc} r_1 & r_2 & r_2\\ 0 & s_2 & 0\\0 & 0 &
s_2 \end{array}\right)$
&
 $\begin{array}{c} a_1=A_1r_1^{-1} \\
 a_2=(A_2r_1-A_1r_2)r_1^{-1}s_2^{-1}\\b_2=B_3s_2^{-1}\end{array}$\\
%&&&&\\
\cline{3-5}
$F^{A2}~=~$&&$\left(\begin{array}{ccc} A_1 & A_2 & A_2\\ 0 & B_2 & 0\\0 & 0 &
B_2 \end{array}\right)$&
$\left(\begin{array}{ccc} r_1 & r_2 & r_2\\ 0 & 0 & s_3\\0 & s_3 &
0 \end{array}\right)$
&
 $\begin{array}{c} a_1=A_1r_1^{-1} \\
 a_2=(A_2r_1-A_1r_2)r_1^{-1}s_3^{-1}\\b_2=B_2s_3^{-1}\end{array}$\\
%&&&&\\
\cline{2-5}
$\left(\begin{array}{ccc} a_1 & a_2 & a_2\\ 0 & 0 & b_2\\0 & b_2 & 0 \end{array}
\right)$&$4-$zero$\times$$6-$zero&$\left(\begin{array}{ccc} A_1 & A_2 & A_2\\ 0 & 0 & B_3\\0 & B_3 &
0 \end{array}\right)$&
$\left(\begin{array}{ccc} r_1 & 0 & 0\\ 0 & s_2 & 0\\0 & 0 &
s_2 \end{array}\right)$
&
 $\begin{array}{c} a_1=A_1r_1^{-1} \\
 a_2=A_2s_2^{-1}\\b_2=B_3s_2^{-1}\end{array}$\\
%&&&&\\
\cline{3-5}
&&$\left(\begin{array}{ccc} A_1 & A_2 & A_2\\ 0 & B_2 & 0\\0 & 0 &
B_2\end{array}\right)$&
$\left(\begin{array}{ccc} r_1 & 0 & 0\\ 0 & 0 & s_3\\0 & s_3 &
0 \end{array}\right)$
&
 $\begin{array}{c} a_1=A_1r_1^{-1} \\
 a_2=A_2s_3^{-1}\\b_2=B_3s_3^{-1}\end{array}$\\
\cline{2-5}
&$6-$zero$\times$$4-$zero&$\left(\begin{array}{ccc} A_1 & 0 & 0\\ 0 & 0 & B_3\\0 & B_3 &
0 \end{array}\right)$&
$\left(\begin{array}{ccc} r_1 & r_2 & r_2\\ 0 & s_2 & 0\\0 & 0 &
s_2 \end{array}\right)$
&
 $\begin{array}{c} a_1=A_1r_1^{-1} \\
 a_2=-A_1r_2r_1^{-1}s_2^{-1}\\b_2=B_2s_2^{-1}\end{array}$\\
%&&&&\\
\cline{3-5}
&&$\left(\begin{array}{ccc} A_1 & 0 & 0\\ 0 & B_2 & 0\\0 & 0 &
B_2 \end{array}\right)$&
$\left(\begin{array}{ccc} r_1 & r_2 & r_2\\ 0 & 0 & s_3\\0 & s_3 &
0 \end{array}\right)$
&
 $\begin{array}{c} a_1=A_1r_1^{-1} \\
 a_2=-A_1r_2r_1^{-1}s_3^{-1}\\b_2=B_3s_3^{-1}\end{array}$\\
\hline
\end{tabular}
%\end{ruledtabular}
\end{table}

%\begin{longtable*}
\begin{table}[!htb]
\caption{\label{tabxtot}  Deconstructed maximal zero textures of $F^{B1}$ and
$F^{B2}$ in category $B$.} 
\begin{tabular}{|c|c|c|c|c|}
\hline
$F$ & Type & $M_D$ & $M$ & Relations\\
\hline
&$4-$zero$\times$$4-$zero&$\left(\begin{array}{ccc} A_1 & 0 & 0\\ B_1 & 0 & B_3\\B_1 & B_3 &
0\end{array}\right)$&
$\left(\begin{array}{ccc} r_1 & 0 & 0\\ s_1 & 0 & s_3\\s_1 & s_3 &
0 \end{array}\right)$
&
 $\begin{array}{c} a_1=A_1r_1^{-1} \\
 b_1=(B_1s_3-B_3s_1)r_1^{-1}s_3^{-1}\\b_2=B_3s_3^{-1}\end{array}$\\
%&&&&\\
\cline{3-5}
$F^{B1}~=$&&$\left(\begin{array}{ccc} A_1 & 0 & 0\\ B_1 & B_2 & 0\\B_1 & 0 &
B_2 \end{array}\right)$&
$\left(\begin{array}{ccc} r_1 & 0 & 0\\ s_1 & s_2 & 0\\s_1 & 0 &
s_2 \end{array}\right)$
&
 $\begin{array}{c} a_1=A_1r_1^{-1} \\
 b_1=(B_1s_2-B_2s_1)r_1^{-1}s_2^{-1}\\b_2=B_2s_2^{-1}\end{array}$\\
%&&&&\\
\cline{2-5}
$\left(\begin{array}{ccc} a_1 & 0 & 0\\ b_1 & b_2 & 0\\b_1 & 0 & b_2  \end{array}
\right)$&$4-$zero$\times$$6-$zero&$\left(\begin{array}{ccc} A_1 & 0 & 0\\ B_1 & 0 & B_3\\B_1 & B_3 &
0 \end{array}\right)$&
$\left(\begin{array}{ccc} r_1 & 0 & 0\\ 0 & 0 & s_3\\0 & s_3 &
0 \end{array}\right)$
&
 $\begin{array}{c} a_1=A_1r_1^{-1} \\
 b_1=B_1r_1^{-1}\\b_2=B_3s_3^{-1}\end{array}$\\
%&&&&\\
\cline{3-5}
&&$\left(\begin{array}{ccc} A_1 & 0 & 0\\ B_1 & B_2 & 0\\B_1 & 0 &
B_2 \end{array}\right)$&
$\left(\begin{array}{ccc} r_1 & 0 & 0\\ 0 & s_2 & 0\\0 & 0 & s_2 \end{array}\right)$
&
 $\begin{array}{c} a_1=A_1r_1^{-1} \\
 b_1=B_1r_1^{-1}\\b_2=B_2s_2^{-1}\end{array}$\\
\cline{2-5}
&$6-$zero$\times$$4-$zero&$\left(\begin{array}{ccc} A_1 & 0 & 0\\ 0 & 0 & B_3\\0 & B_3 &
0 \end{array}\right)$&
$\left(\begin{array}{ccc} r_1 & 0 & 0\\ s_1 & 0 & s_3\\s_1 & s_3 &
0 \end{array}\right)$
&
 $\begin{array}{c} a_1=A_1r_1^{-1} \\
 b_1=-B_3s_1r_1^{-1}s_3^{-1}\\b_2=B_3s_3^{-1}\end{array}$\\
%&&&&\\
\cline{3-5}
&&$\left(\begin{array}{ccc} A_1 & 0 & 0\\ 0 & B_2 & 0\\0 & 0 &
B_2 \end{array}\right)$&
$\left(\begin{array}{ccc} r_1 & 0 & 0\\ s_1 & s_2 & 0\\s_1 & 0 & s_2 \end{array}\right)$
&
 $\begin{array}{c} a_1=A_1r_1^{-1} \\
b_1=-B_2s_1r_1^{-1}s_2^{-1}\\b_2=B_2s_2^{-1}\end{array}$\\
%\hline
%\end{tabular}
%\end{ruledtabular}
%\end{table}
%\begin{longtable*}
%\begin{table}[!ht]
%\caption{\label{tabxtot}  Deconstructed maximal zero textures for $F^{B2}$.} 
%\begin{tabular}{|c|c|c|c|c|}
%\hline
%$F^{B2}$ & Type & $m_D$ & $M$ & Relations\\
\hline
&$4-$zero$\times$$4-$zero&$\left(\begin{array}{ccc} A_1 & 0 & 0\\ B_1 & 0 & B_3\\B_1 & B_3 &
0\end{array}\right)$&
$\left(\begin{array}{ccc} r_1 & 0 & 0\\ s_1 & s_2 & 0\\s_1 & 0 &
s_2 \end{array}\right)$
&
 $\begin{array}{c} a_1=A_1r_1^{-1} \\
 b_1=(B_1s_2-B_3s_1)r_1^{-1}s_2^{-1}\\b_2=B_3s_2^{-1}\end{array}$\\
%&&&&\\
\cline{3-5}
$F^{B2}~=$&&$\left(\begin{array}{ccc} A_1 & 0 & 0\\ B_1 & B_2 & 0\\B_1 & 0 &
B_2 \end{array}\right)$&
$\left(\begin{array}{ccc} r_1 & 0 & 0\\ s_1 & 0 & s_3\\s_1 & s_2 &
0 \end{array}\right)$
&
 $\begin{array}{c} a_1=A_1r_1^{-1} \\
 b_1=(B_1s_3-B_2s_1)r_1^{-1}s_3^{-1}\\b_2=B_2s_3^{-1}\end{array}$\\
%&&&&\\
\cline{2-5}
$\left(\begin{array}{ccc} a_1 & 0 & 0\\ b_1 & 0 & b_2\\b_1 & b_2 & 0 \end{array}
\right)$&$4-$zero$\times$$6-$zero&$\left(\begin{array}{ccc} A_1 & 0 & 0\\ B_1 & 0 & B_3\\B_1 & B_3 &
0 \end{array}\right)$&
$\left(\begin{array}{ccc} r_1 & 0 & 0\\ 0 & s_2 & 0\\0 & 0 &
s_2 \end{array}\right)$
&
 $\begin{array}{c} a_1=A_1r_1^{-1} \\
 b_1=B_1r_1^{-1}\\b_2=B_3s_2^{-1}\end{array}$\\
%&&&&\\
\cline{3-5}
&&$\left(\begin{array}{ccc} A_1 & 0 & 0\\ B_1 & B_2 & 0\\B_1 & 0 &
B_2 \end{array}\right)$&
$\left(\begin{array}{ccc} r_1 & 0 & 0\\ 0 & 0 & s_3\\0 & s_3 & 0 \end{array}\right)$
&
 $\begin{array}{c} a_1=A_1r_1^{-1} \\
 b_1=B_1r_1^{-1}\\b_2=B_2s_3^{-1}\end{array}$\\
\cline{2-5}
&$6-$zero$\times$$4-$zero&$\left(\begin{array}{ccc} A_1 & 0 & 0\\ 0 & 0 & B_3\\0 & B_3 &
0 \end{array}\right)$&
$\left(\begin{array}{ccc} r_1 & 0 & 0\\ s_1 & s_2 & 0\\s_1 & 0 &
s_2 \end{array}\right)$
&
 $\begin{array}{c} a_1=A_1r_1^{-1} \\
 b_1=-B_3s_1r_1^{-1}s_2^{-1}\\b_2=B_3s_2^{-1}\end{array}$\\
%&&&&\\
\cline{3-5}
&&$\left(\begin{array}{ccc} A_1 & 0 & 0\\ 0 & B_2 & 0\\0 & 0 &
B_2 \end{array}\right)$&
$\left(\begin{array}{ccc} r_1 & 0 & 0\\ s_1 & 0 & s_3\\s_1 & s_3 & 0 \end{array}\right)$
&
 $\begin{array}{c} a_1=A_1r_1^{-1} \\
b_1=-B_2s_1r_1^{-1}s_3^{-1}\\b_2=B_2s_3^{-1}\end{array}$\\
\hline
\end{tabular}
%\end{ruledtabular}
\end{table}
\section{Constraints from the requirement of leptogenesis}
For the inverse seesaw, Dirac leptogenesis \cite{six} 
occurs at an energy scale where the  $\mu$ term,  required 
to be
$O(\rm{keV})$, can be neglected. Gauge symmetry is preserved and 
lepton number is conserved at that stage. 
The relevant part of the Lagrangian for
leptogenesis (LG) is
\begin{eqnarray}
-{\cal L}^{\rm
 LG}=\frac{\sqrt
 2}{v}\overline{\psi_{kL}}(M_D)_{kk'}{\tilde \Phi}\nu_{k'R}+\overline{s_{kL}}M_{kk'}\nu_{k'R}
 + h.c., 
\label{yuklag}
\end{eqnarray}
where $\psi_L\equiv\left(\begin{array}{c}\nu_L\\ l \end{array}\right)$ and
$ \Phi\equiv\left(\begin{array}{c}\phi^+\\ \phi^0 \end{array}\right)$ are 
weak isospin doublet lepton and Higgs fields respectively and 
$v \approx 246$ GeV.
With the definition of a Dirac spinor field 
$N\equiv s_L \bigoplus \nu_R$, we have 
\begin{eqnarray}
-{\cal L}^{LG}
 =\frac{\sqrt
 2}{v}\overline{\psi_{kL}}(M_D)_{kk'}{\tilde \Phi}N_{k'R}+\overline{N_{kL}}M_{kk'}N_{k'R}
 + h.c.. 
\label{yuklag2}
\end{eqnarray}
\par
The CP asymmetry in the  decays of $N$ and 
${\bar N}$ is characterised in their mass basis by the parameter
$\varepsilon_{il}$ defined as 
\begin{eqnarray}
\varepsilon_{il}\equiv\frac{\Gamma ({N_{i}\rightarrow
l^-\phi^+,\nu_l\phi^0})-\Gamma ({{\bar N_{i}}\rightarrow l^+\phi^-,
\nu_l^C\phi^{0*}})}{\Gamma({N_{i}\rightarrow
l^-\phi^+,\nu_l\phi^0})+\Gamma({{\bar N_{i}}
\rightarrow l^+\phi^-,\nu_l^C\phi^{0*}})}.
\label{lepasym}
\end{eqnarray}
The singlet neutrino Dirac mass matrix $M$ can be put into a diagonal form 
by a biunitary transformation 
\begin{eqnarray}
U_L^\dagger M U_R={\rm diag}\,(M_1,~M_2,~M_3).
\label{mrd}
\end{eqnarray}
The important unitary matrix for leptogenesis is $U_R$ which diagonalises 
$M^\dagger M$: 
\begin{eqnarray}
U_R^\dagger M^\dagger M U_R={\rm diag}\,(M_1^2,~M_2^2,~M_3^2).
\label{hrd}
\end{eqnarray}
In the mass basis of $N$, where it is denoted by a 
widehat, 
$\widehat{N_{iR}}=(U_R^\dagger)_{il}N_{lR}$, i.e. 
$N_{lR}=(U_R)_{li}{\widehat{N_{iR}}}$. In this basis of $N$, $M_D$ 
is modified to 
\begin{eqnarray}
\widehat{M_D}=M_D U_R.  
\label{mdfmd}
\end{eqnarray}
On account of the Dirac nature of the mass matrix 
$M$, the CP asymmetry, as generated from the 
interference of the tree and one
loop self energy diagrams, will be 
\begin{eqnarray}
\varepsilon_{il}=\frac{M_i^2}{4\pi v^2 \widehat{h_{ii}}} 
\sum_j\frac{M_i^2-M_j^2}{(M_i^2-M_j^2)^2
+\Gamma_i^2M_i^2} {\rm Im}[\widehat{M_{D lj}}\widehat{M_{Dli}^*} 
\widehat{h}_{ji}],
\label{lepasym1}
\end{eqnarray}
where $\widehat{h} = \widehat{M_D^\dagger} \widehat{M_D}=  
U_R^\dagger M_D^\dagger M_D  
U_R\equiv U_R^\dagger h U_R$ and $\Gamma_i$ is the total width of $N_i$.
\par
There are four allowed 4-zero and two allowed 6-zero textures of
$M$ :  
\begin{eqnarray}
\left(\begin{array}{ccc} r_1 & r_2 & r_2\\ 0 & s_2 & 0\\0 & 0 &
s_2 \end{array}\right),\qquad \left(\begin{array}{ccc} r_1 & r_2 & r_2\\ 0 & 0 & s_2\\0 & s_2 &
0 \end{array}\right),\nonumber\\
\left(\begin{array}{ccc} r_1 & 0 & 0\\ s_1 & s_2 & 0\\s_1 & 0 &
s_2 \end{array}\right), \qquad \left(\begin{array}{ccc} r_1 & 0 & 0\\ s_1 & 0 &
s_2\\s_1 & s_2 & 0 \end{array}\right);\nonumber\\
\left(\begin{array}{ccc} r_1 & 0 & 0\\ 0 & s_2 & 0\\0 & 0 &
s_2 \end{array}\right), \qquad \left(\begin{array}{ccc} r_1 & 0 & 0\\ 0 & 0 & s_2\\0 & s_2 & 0 \end{array}\right).
\end{eqnarray}
\noindent
Each pair of $M$  occuring in the three lines of the 
above equation leads to the following
three respective forms for the matrix $H\equiv M^\dagger M$ :
\begin{eqnarray}
H = &&\left(\begin{array}{ccc} |r_1|^2 & r_1^*r_2 & r_1^*r_2\\ r_1r_2^* & |r_2|^2+|s_2|^2 & |r_2|^2\\r_1r_2^* & |r_2|^2 &
|r_2|^2+|s_2|^2 \end{array}\right),\nonumber\\
H = &&~~~\left(\begin{array}{ccc} |r_1|^2+2|r_2|^2 & r_2^*s_2 & r_2^*s_2\\ r_2s_2^* &
|s_2|^2  & 0\\r_2s_2^* & 0 &
|s_2|^2 \end{array}\right),\nonumber\\
H = &&~~~~~~~~\left(\begin{array}{ccc} |r_1|^2 & 0 & 0\\ 0 &
|s_2|^2 & 0\\ 0 & 0 &
|s_2|^2 \end{array}\right).
\label{HH}
\end{eqnarray}
\noindent
The top two $H$ matrices of (\ref{HH}) can be diagonalised by 
$U_R$ which can be defined  by one 
unknown angle $\phi_{12}$ and one phase $\psi$ :
\begin{eqnarray}
U_R = \left(\begin{array}{ccc} c_{12}^\phi e^{-i\psi} &
s_{12}^\phi e^{-i\psi} & 0\\ -\frac{s_{12}^\phi}{\sqrt 2} & \frac{c_{12}^\phi}{\sqrt 2}
& \frac{-1}{\sqrt 2}\\ \frac{-s_{12}^\phi}{\sqrt 2} & \frac{c_{12}^\phi}{\sqrt 2} &
\frac{1}{\sqrt 2} \end{array}\right),
\end{eqnarray}
where $c_{12}^\phi\equiv\cos\phi_{12},~s_{12}^\phi\equiv\sin\phi_{12}$. 
The mass eigenvalues 
$M_i$ as well as the angle 
$\phi_{12}$ and the phase angle $\psi$ of $U_R$ are given in Table 3 for the 
two different pairs 
of 4-zero $M$'s. The $H$ matrix corresponding to the $6$-zero textures is
diagonal. Hence, for them,  
the diagonalising matrix will be $U_R={\bf I}$ and the
eigenvalues of $H$ will be the diagonal elements 
$M_1^2=|r_1|^2,~M_{2,3}^2=|s_2|^2$. 
%%%%%%%%%%%%%%%%%%%%%%%%%%%%%%%%%%%%%%%%%%%%%%%%%%%%%
%\begin{longtable*}
\begin{table}[!htb]
\caption{\label{tabxtot} Mass eigenvalues, $\psi$ and $\phi_{12}$ for two 
categories of $H$ from 
4-zero $M$'s.} 
\begin{tabular}{|c|c|c|}
\hline
 $M^\dagger M$  & $M_i^2$ & $\psi$ and $\phi_{12}$\\
\hline
$\left(\begin{array}{ccc} |r_1|^2 & r_1^*r_2 & r_1^*r_2\\ r_1r_2^* & |r_2|^2+|s_2|^2 & |r_2|^2\\r_1r_2^* & |r_2|^2 &
|r_2|^2+|s_2|^2 \end{array}\right)$&
 $\begin{array}{c}
 M^2_{2,1}=\frac{|r_1|^2+2|r_2|^2+|s_2|^2}{2}\\
\pm{
 [{(|s_2|^2+2|r_2|^2-|r_1|^2)}^2+8|r_1||r_2|]}^{1/2}\\
M_3^2=|s_2|^2 \end{array}$&
$\begin{array}{c} \psi ={\rm arg
}(r_1r_2^*)\\ 
\tan\phi_{12}=\frac{\sqrt{2|r_1r_2|}}{M_2^2 -|r_1|^2}\end{array}$\\
\hline
$\left(\begin{array}{ccc} |r_1|^2+2|r_2|^2 & r_2^*s_2 & r_2^*s_2\\ r_2s_2^* &
|s_2|^2  & 0\\r_2s_2^* & 0 &
|s_2|^2 \end{array}\right)$&$\begin{array}{c}
 M^2_{2,1}=\frac{|r_1|^2+2|r_2|^2+|s_2|^2}{2}\\
\pm{
 [{(|r_1|^2+2|r_2|^2-|s_2|^2)}^2+8|r_2||s_2|]}^{1/2}\\M_3^2=|s_2|^2 \end{array}$&$\begin{array}{c} \psi ={\rm arg
}(r_2s_2^*)\\ \tan\phi_{12}=\frac{{\sqrt 2}|r_2s_2^*|}{M_2^2 -|r_1|^2-2r_2^2}\end{array}$\\
\hline
\end{tabular}
%\end{ruledtabular}
\end{table}

Now that we have the $U_R$'s for all possible $M^\dagger M$ matrices, 
we can construct the
matrices $\widehat{M_D} = M_DU_R$ and $\widehat{h}$, which are  relevant to 
leptogenesis, 
for all possible combinations of $M_D$ and $M$ given in Tables 1 and 2. 
We mainly focus on the 
determination of ${\rm Im} 
\widehat{(M_{Dlj}} \widehat{M_{Dli}^*} \widehat{h_{ji}})$, 
cf. (\ref{lepasym1}), 
and test whether this quantity is nonzero. We observe that, 
for each $\mu\tau$ symmetric $4$-zero $\times$
$6$-zero combination, $U_R={\bf I}$, $\widehat{M_D}=M_D$, $\widehat{h}=h$ 
and the said quantity vanishes. 
The same is true for all $\mu\tau$ symmetric $6$-zero $\times$
$4$-zero combinations. So, not one of the allowed texures of the
 $4$-zero $\times$
$6$-zero and $6$-zero $\times$
$4$-zero combinations is able to create a nonvanishing 
lepton asymmetry which we require. Hence all these texture combinations 
are ruled out. Only the allowed $4$-zero$\, \times$
\,$4$-zero combinations lead to 
$\varepsilon_{il}\ne0$ for $i=1,~2$ and all $l$. They are just 
the ones to survive the requirement of leptogenesis.
%%%%%%%%%%%%%%%%%%%%%%%%%%%%%%%%%%%%%%%%%%%%
\section{Effect of $\mu\tau$ symmetry breaking on the surviving textures}
The symmetry, arising from $\mu\tau$ interchange, produces a 
maximal atmospheric mixing angle $\theta_{23}=\pi/4$ as well as a 
vanishing reactor angle  $\theta_{13}=0$.
 In the light of the recent results of reactor and short baseline 
experiments measuring $\theta_{13}\simeq 8^o$, it is clear the 
$\mu\tau$ symmetry is broken. The scale of this breaking can be 
estimated in terms of parameters 
$\{\epsilon\}\sim 15{\%}$ from the magnitude of the dimensionless quantity 
$\sin\theta_{13}\simeq 0.15$.
We realise the breaking of $\mu\tau$ symmetry in $M_D$ and 
in $\mu$ for the surviving $4$-zero$\times$ $4$-zero combinations of $M_D$ 
and $M$ in terms of two complex parameters 
$\epsilon_1e^{i\phi_1}$, $\epsilon_2e^{i\phi_2}$ where $\epsilon_{1,2}$ and 
$\phi_{1,2}$ are real. 
We assume that the TeV scale $M$ remains unbroken in a 
$\mu\tau$ symmetric form. In addition, we keep the positions of zeros in
 $M_D$, $M$ and $\mu$ in tact. This is since we do not want to change 
the texture pattern. In this set up, $\mu\tau$ symmetry is broken in the diagonal matrix $\mu$ by means of a 
parameter $\delta$ introduced via 
\begin{eqnarray}
\mu^{\delta}={\rm diag}\left(\mu_1,~\mu_2,~\mu_2(1+\delta)\right).
\label{dmlmu}
\end{eqnarray} 
\noindent
The four allowed forms of $M_D$, now designated $M_D^{\epsilon_1,\epsilon_2}$, 
look like  
\begin{eqnarray}
&&\left(\begin{array}{ccc} A_1 & A_2 & A_2(1+\epsilon_1 e^{i\phi_1})\\ 0 & B_2(1+\epsilon_2 e^{i\phi_2}) & 0\\0 & 0 & B_2 \end{array}
\right),\qquad\qquad\quad
\left(\begin{array}{ccc} A_1 & A_2 & A_2(1+\epsilon_1 e^{i\phi_1})\\ 0 & 0 & B_3(1+\epsilon_2 e^{i\phi_2})\\0 & B_3 & 0 \end{array}
\right),\nonumber\\
&&\left(\begin{array}{ccc} A_1 & 0 & 0\\ B_1(1+\epsilon_1 e^{i\phi_1}) & B_2(1+\epsilon_2 e^{i\phi_2}) & 0\\B_1 & 0 & B_2  \end{array}
\right),\quad
\left(\begin{array}{ccc} A_1 & 0 & 0\\ B_1(1+\epsilon_1 e^{i\phi_1}) & 0 & B_3(1+\epsilon_2 e^{i\phi_2})\\B_1 & B_3 & 0 \end{array}
\right).\nonumber\\
\label{al40}
\end{eqnarray}
\par
The top two of the four matrices in (\ref{al40}) occur in the 
$4$-zero $\times$ $4$-zero parts of Category $A$ in Table 1. The bottom two 
matrices above occur in the    
$4$-zero $\times$ $4$-zero parts of Category $B$ in Table 2. 
Using  the inverse see-saw formula (\ref{invmas})
with broken $\mu\tau$ symmetric mass matrices 
$\mu^\delta$, $M_D^{\epsilon_1,\epsilon_2}$ 
and
 an unbroken $M$, we have found four  distinct forms of 
$M_\nu^{\epsilon_1,\epsilon_2\delta}$ 
for the four $4$-zero $\times$ $4$-zero combinations of category $A$ as given in (\ref{mnu1})-(\ref{mnu4})
Those combinations, each with a broken $\mu\tau$ symmetric $M_D$ and a 
$\mu\tau$ 
symmetric $M$, are named as $A1$, $A2$, $A3$ and $A4$ in respective order.
 We have used three real parameters $k_1$,
 $k_2$ and $k_3$, three phases $e^{i\alpha}$, 
$e^{i\beta}$ and $e^{i\gamma}$, and an overall complex mass parameter
 $m_0$. The definitions of these quantities are given in Table 4. 
\begin{table}[!htb]
\begin{center}
\caption{\label{tabxtot} Definitions of parameters used in mass matrices for different categories.} 
\begin{tabular}{|c|c|c|c|c|}
\hline
 Parameters  &
\multicolumn{4}{c|}{Definition of Parameters for Category}\\
\cline{2-5}
&A1&A2&A3&A4\\
\hline
$m_0$&${B_2^2\mu_2}/{s_2^2}$&${B_3^2\mu_2}/{s_3^2}$&${B_3^2\mu_2}/{s_2^2}$&$
{B_2^2\mu_2}/{s_3^2}$\\
\hline
$k_1e^{i(\alpha+\gamma)}$& ${A_1s_2}\sqrt{{\mu_1}/{\mu_2}}/B_2r_1$&
${A_1s_3}\sqrt{{\mu_1}/{\mu_2}}/B_3r_1$&${A_1s_2}\sqrt{{\mu_1}/
{\mu_2}}/B_3r_1$&
${A_1s_3}/\sqrt{{\mu_1}/{\mu_2}}/B_2r_1$\\
\hline
$k_2e^{i(\beta+\gamma)}$&${A_2}/{B_2}$&${A_2}/{B_3}$&${A_2}/{B_3}$&${A_2}/{B_2}$\\
\hline
$k_3e^{i\gamma}$&${A_1r_2}/{B_2r_1}$&${A_1r_2}/{B_3r_1}$&${A_1r_2}/{B_3r_1}$&${A_1r_2}/{B_2r_1}$\\
\hline
\hline
%\cline{2-5}
&B1&B2&B3&B4\\
\cline{2-5}
$m_0$&${B_3^2\mu_2}/{s_3^2}$&${B_2^2\mu_2}/{s_2^2}$&${B_3^2\mu_2}/{s_2^2}$&${B_2^2\mu_2}/{s_3^2}$\\
\hline
$k_1e^{i\alpha}$& ${B_1s_3}/{B_3r_1}$&
${B_1s_2}/{B_2r_1}$&${B_1s_2}/{B_3r_1}$&
${B_1s_3}/{B_2r_1}$\\
\hline
$k_2e^{i\beta}$&${r_2}/{r_1}$&${r_2}/{r_1}$&${r_2}/{r_1}$&${r_2}/{r_1}$\\
\hline
$k_3e^{i\gamma}$&${A_1s_3}\sqrt{{\mu_1}/{\mu_2}}/B_3r_1$&
${A_1s_2}\sqrt{{\mu_1}/{\mu_2}}/B_2r_1$&${A_1s_2}\sqrt{{\mu_1}/{\mu_2}}/
B_3r_1$&
${A_1s_3}\sqrt{{\mu_1}/{\mu_2}}/B_2r_1$\\
\hline
%\cline{2-5}
%&A1a&A1b&A1c&A1d\\
\end{tabular}
%\end{ruledtabular}
\end{center}
\end{table}
We have also been able to remove phase factor 
$e^{i\gamma}$ from the neutrino mass matrices of all categories
by redefining the neutrino field $\nu_e$.
For Category A, then, we have four forms of 
${M_\nu}^{\epsilon_1,\epsilon_2,\delta}$ 
as follows : 
%%%%%%1
\begin{eqnarray}
 m_0\left(\begin{array}{ccc}\left[\begin{array}{c}k_1^2e^{2i\alpha}+(k_2e^{i\beta} 
-k_3)^2\\
+\{k_2e^{i\beta}(1+\epsilon_1e^{i\phi_1})-k_3\}^2(1+\delta)\end{array}\right]& 
 (k_2e^{i\beta}-k_3)(1+\epsilon_2e^{i\phi_2}) &
\{k_2e^{i\beta}(1+\epsilon_1e^{i\phi_1})-k_3\}(1+\delta)\\
(k_2e^{i\beta}-k_3)(1+\epsilon_2e^{i\phi_2})&(1+\epsilon_2e^{i\phi_2})^2&0\\
(k_2e^{i\beta}(1+\epsilon_1e^{i\phi_1})-k_3)(1+\delta)&0&1+\delta
 \end{array}\right),\nonumber\\
\label{mnu1}
\end{eqnarray}
%
%2
%%%%%%
\begin{eqnarray}
 m_0\left(\begin{array}{ccc}\left[\begin{array}{c}k_1^2e^{2i\alpha}+(k_2e^{i\beta} 
-k_3)^2(1+\delta)\\
+\{k_2e^{i\beta}(1+\epsilon_1e^{i\phi_1})-k_3\}^2\end{array}\right]& 
 \{k_2e^{i\beta}(1+\epsilon_1e^{i\phi_1})-k_3\}(1+\epsilon_2e^{i\phi_2}) &
(k_2e^{i\beta}-k_3)(1+\delta)\\
\{k_2e^{i\beta}(1+\epsilon_1e^{i\phi_1})-k_3\}
(1+\epsilon_2e^{i\phi_2})&(1+\epsilon_2e^{i\phi_2})^2&0\\
(k_2e^{i\beta}-k_3)(1+\delta)&0&1+\delta
 \end{array}\right),\nonumber\\
\end{eqnarray}
%
%3
%%%%%%
\begin{eqnarray}
 m_0\left(\begin{array}{ccc}\left[\begin{array}{c}k_1^2e^{2i\alpha}+(k_2e^{i\beta} 
-k_3)^2\\
+\{k_2e^{i\beta}(1+\epsilon_1e^{i\phi_1})-k_3\}^2(1+\delta)\end{array}\right]& 
 \{k_2e^{i\beta}(1+\epsilon_1e^{i\phi_1})-k_3\}(1+\epsilon_2e^{i\phi_2})(1+\delta) &
k_2e^{i\beta}-k_3\\
\{k_2e^{i\beta}(1+\epsilon_1e^{i\phi_1})-k_3\}
(1+\epsilon_2e^{i\phi_2})(1+\delta)&(1+\delta)(1+\epsilon_2e^{i\phi_2})^2&0\\
k_2e^{i\beta}-k_3&0&1
 \end{array}\right),\nonumber\\
\end{eqnarray}
and
%%%%%%4
\begin{eqnarray}
 m_0\left(\begin{array}{ccc}\left[\begin{array}{c}k_1^2e^{2i\alpha}+(k_2e^{i\beta} 
-k_3)^2(1+\delta)\\
+\{k_2e^{i\beta}(1+\epsilon_1e^{i\phi_1})-k_3\}^2\end{array}\right]& 
 (k_2e^{i\beta}-k_3)(1+\epsilon_2e^{i\phi_2})(1+\delta) &
k_2e^{i\beta}(1+\epsilon_1e^{i\phi_1})-k_3\\
(k_2e^{i\beta}-k_3)(1+\epsilon_2e^{i\phi_2})(1+\delta)&(1+\delta)(1+\epsilon_2e^{i\phi_2})^2&0\\
k_2e^{i\beta}(1+\epsilon_1e^{i\phi_1})-k_3&0&1
 \end{array}\right).\nonumber\\
\label{mnu4}
\end{eqnarray}
For category $B$, four combinations of a broken $\mu\tau$ symmetric 
$M_D$ and $\mu\tau$ 
symmetric $M$ are named as as $B1$, $B2$, $B3$ and $B4$ in respective order. 
$B1$ and $B2$ 
 lead to one form of
$M_\nu^{\epsilon_1,\epsilon_2,\delta}$ whereas $B3$ and $B4$ to
 another form of 
$M_\nu^{\epsilon_1,\epsilon_2,\delta}$. Those  
are
 given below respectively as
%%%%%%1
\begin{eqnarray}
 m_0\left(\begin{array}{ccc}k_3^2& 
       \left[\begin{array}{c}k_3k_1e^{i\alpha}(1+\epsilon_1e^{i\phi_1})\\ 
-k_3k_2e^{i\beta}\end{array}\right] &k_3\{k_1e^{i\alpha}-k_2e^{i\beta}\}\\
 \left[\begin{array}{c}k_3k_1e^{i\alpha}(1+\epsilon_1e^{i\phi_1})\\ 
-k_3k_2e^{i\beta}\end{array}\right]&\left[\begin{array}{c}\left\{k_1e^{i\alpha}(1+\epsilon_1e^{i\phi_1})
\right.\\ \left.
-k_2e^{i\beta}(1+\epsilon_2e^{i\phi_2})\right\}^2\\+(1+\epsilon_2e^{i\phi_1})^2\end{array}\right]&
\left[\begin{array}{c}\left\{k_1e^{i\alpha}(1+\epsilon_1e^{i\phi_1})\right.\\\left.
-k_2e^{i\beta}(1+\epsilon_2e^{i\phi_2})\right\}\\ \times \{k_1e^{i\alpha}-k_2e^{i\beta}\}\end{array}\right]\\
k_3\{k_1e^{i\alpha}-k_2e^{i\beta}\}&\left[\begin{array}{c}\left\{k_1e^{i\alpha}(1+\epsilon_1e^{i\phi_1})\right.\\\left.
-k_2e^{i\beta}(1+\epsilon_2e^{i\phi_2})\right\}\\ \times \{k_1e^{i\alpha}-k_2e^{i\beta}\}\end{array}\right]&
\{k_1e^{i\alpha}-k_2e^{i\beta}\}^2+1+\delta
  \end{array}\right),\nonumber\\
\end{eqnarray}
%%%%%%2
\begin{eqnarray}
 m_0\left(\begin{array}{ccc}k_3^2& 
       \left[\begin{array}{c}k_3k_1e^{i\alpha}(1+\epsilon_1e^{i\phi_1})\\ 
-k_3k_2e^{i\beta}\end{array}\right] &k_3\{k_1e^{i\alpha}-k_2e^{i\beta}\}\\
 \left[\begin{array}{c}k_3k_1e^{i\alpha}(1+\epsilon_1e^{i\phi_1})\\ 
-k_3k_2e^{i\beta}\end{array}\right]&\left[\begin{array}{c}\left\{k_1e^{i\alpha}(1+\epsilon_1e^{i\phi_1})
\right.\\ \left.
-k_2e^{i\beta}(1+\epsilon_2e^{i\phi_2})\right\}^2\\+(1+\delta)(1+\epsilon_2e^{i\phi_1})^2\end{array}\right]&
\left[\begin{array}{c}\left\{k_1e^{i\alpha}(1+\epsilon_1e^{i\phi_1})\right.\\\left.
-k_2e^{i\beta}(1+\epsilon_2e^{i\phi_2})\right\}\\ \times \{k_1e^{i\alpha}-k_2e^{i\beta}\}\end{array}\right]\\
k_3\{k_1e^{i\alpha}-k_2e^{i\beta}\}&\left[\begin{array}{c}\left\{k_1e^{i\alpha}(1+\epsilon_1e^{i\phi_1})\right.\\\left.
-k_2e^{i\beta}(1+\epsilon_2e^{i\phi_2})\right\}\\ \times \{k_1e^{i\alpha}-k_2e^{i\beta}\}\end{array}\right]&
\{k_1e^{i\alpha}-k_2e^{i\beta}\}^2+1
  \end{array}\right).\nonumber\\
\label{mnu6}
\end{eqnarray}
So far, then, there are six distinct allowed forms of $M_\nu^{\epsilon_1,\epsilon_2,\delta}$
\par
It is to be pointed out, that apart from $4$-zero $\times$ $4$-zero 
combinations 
we can break $\mu\tau$ symmetry also 
 in  the $4$-zero $\times$ $6$-zero and $6$-zero $\times$ $4$-zero 
combinations. A 
question may then arise. Though the
 latter combinations with $\mu\tau$ symmetry are ruled out from 
the leptogenesis requirement, are they relevant
after $\mu\tau$ symmetry breaking? We have observed that, for the 
$4$-zero $\times$ $6$-zero combinations, $M^\dagger M$ is
diagonal. So,
$M_D^{\epsilon_1,\epsilon_2}$ retains the  same four zero structure 
in the mass basis of $N$ as in the $\mu\tau$ symmetric cases.
 Hence, the lepton asymmetry still vanishes for those combinations even 
after $\mu\tau$ symmetry breaking. For broken $\mu\tau$
 symmetric $6$-zero $\times$ $4$-zero combinations, we have again calculated
${\rm Im} 
\widehat{(M_{Dlj}} \widehat{M_{Dli}^*} \widehat{h_{ji}})$ and have found 
a vanishing lepton asymmetry in each case.
So, given the requirement of leptogenesis, only the 
$4$-zero $\times$ $4$-zero 
combinations are left for study
even after $\mu\tau$ symmetry breaking.

\section{Exact diagonalisation of the neutrino mass matrix}
We have written all neutrino mass matrices in the full 
fledged form without any approximation. So, we can perform 
an exact analysis by diagonalising them. Recently,
an exact diagonalisation method was introduced \cite{Adhikary:2013bma}
to find out the masses, mixing angles and phases in terms of the 
neutrino mass matrix elements. We use that methodology next 
to determine the physical observables. We just present the results here, 
for details 
one can see \cite{Adhikary:2013bma}.
\par
In order to find out the squared masses, mixing angles and the 
Dirac CP phase, we have 
constructed the matrices  
$h=M_\nu^{\epsilon_1,\epsilon_2,\delta}
(M_\nu^{\epsilon_1,\epsilon_2,\delta})^\dagger$
for all the above six neutrino mass matrices and have diagonalised them. 
The general form of the eigenvalues of $h$ are
\begin{eqnarray}
&&\lambda_1=-\frac{b}{3a}-\frac{2\sqrt[3]{r}}{3\sqrt[3]{2}a}\cos \theta ,\nonumber\\
&&\lambda_2=-\frac{b}{3a}+\frac{\sqrt[3]{r}}{3\sqrt[3]{2}a}(\cos \theta -\sqrt{3}\sin \theta),\nonumber\\
&&\lambda_3=-\frac{b}{3a}+\frac{\sqrt[3]{r}}{3\sqrt[3]{2}a}(\cos \theta +\sqrt{3}\sin \theta).
\end{eqnarray} 
where the definitions of $a$, $b$, $r$ and $\theta$ in terms of the 
elements of  
$h$ are
 given in the Appendix. Now the identification of 
$\lambda_1$, $\lambda_2$ and $\lambda_3$ with the 
squared masses $m_i^2$ is
done in order to compare with experimental data.
To determine the mixing angles, we have used the eigenvectors for the 
eigenvalues $m_i^2$ and have 
constructed the diagonalising matrix in terms of the elements of  $h$. 
The elements of the 
diagonalising $U$ matrix can be written as
\begin{eqnarray}
&&U_{1i}=\frac{(h_{22}-m_i^2)h_{13}-h_{12}h_{23}}{N_i},\nonumber\\%=\frac{P_{1i}}{N_i}\\\nonumber\\
&&U_{2i}=\frac{(h_{11}-m_i^2)h_{23}-
{h_{12}}^\ast h_{13}}{N_i},\nonumber\\%=\frac{P_{2i}}{N_i}\\\nonumber\\
&&U_{3i}=\frac{|h_{12}|^2-(h_{11}-
m_i^2)(h_{22}-m_i^2)}{N_i},%=\frac{P_{3i}}{N_i}\label{uij} .
\end{eqnarray}
with i=1,2,3. The $N_i$'s are normalization constants given by
\begin{eqnarray} 
|N_i|^2&=&|(h_{22}-m_i^2)h_{13}-
h_{12}h_{23}|^2+\nonumber\\
&&|(h_{11}-m_i^2)h_{23}-
{h_{12}}^\ast h_{13}|^2+\nonumber\\
&&\{|h_{12}|^2-(h_{11}-
m_i^2)(h_{22}-m_i^2)\}^2 .
\end{eqnarray} 
The elements of the $U$ matrix can have unwanted overall phases 
but their moduli can be equated to the moduli of 
the elements of the mixing matrix (without Majorana phases) : 
\begin{equation}
U^{\rm MIXING}= \left(\begin{array}{ccc}c_{12} c_{13}&
                      s_{12} c_{13}&
                      s_{13} e^{-i\delta_{CP}}\\
-s_{12} c_{23}-c_{12} s_{23} s_{13} e^{i\delta_{CP}}& c_{12} c_{23}-
s_{12} s_{23} s_{13} e^{i\delta_{CP}}&
s_{23} c_{13}\\
s_{12} s_{23} -c_{12} c_{23} s_{13} e^{i\delta_{CP}}&
-c_{12} s_{23} -s_{12} c_{23} s_{13} e^{i\delta_{CP}}&
c_{23} c_{13}\end{array}\right)
\end{equation}
with $c_{ij}\equiv \cos\theta_{ij}$, $s_{ij}\equiv \sin\theta_{ij}$ 
and $\delta_{CP}$ 
being the Dirac phase.
 So, using $|U^{\rm MIXING}_{ij}|=|U_{ij}|$, we have the  expressions for the 
three mixing angles as
\begin{eqnarray}
\tan \theta_{23}=\frac{|U_{23}|}{|U_{33}|},\\\nonumber\\
\tan \theta_{12}=\frac{|U_{12}|}{|U_{11}|},\\\nonumber\\
\sin \theta_{13}=|U_{13}|\label{last}.
\end{eqnarray}
The $\delta_{CP}$ phase can be obtained by using the expression 
for $h_{12}h_{23}h_{31}$: 
\begin{eqnarray}
{\rm Im}(h_{12}h_{23}h_{31})=\frac{1}{8}{(m_2^2-m_1^2)(m_3^2-m_2^2)
(m_3^2-m_1^2)\sin2\theta_{12}\sin2\theta_{23}\sin2\theta_{13}
\cos\theta_{13}\sin\delta_{CP}}.\nonumber\\
\label{cp}
\end{eqnarray}
\par
Eqn.(\ref{cp}) can be inverted to obtain the Dirac CP phase.
An exact knowledge of the masses, mixing angles and the Dirac CP phase 
in terms of the mass elements of the neutrino mass 
matrix enables us to obtain expressions for the Majorana phases,  
neglecting terms with $(c_{23}^2-s_{23}^2)s_{13}$, $s_{13}^2$ and 
their higher powers. Thus we can write
\begin{eqnarray}
  \tan{\theta_j}=\frac{Y'_jW_j-W'_jY_j}{X_jW'_j-W_jX'_j},
 \end{eqnarray}
where $j=1,2$ and $\theta_1=\alpha_M$,  $\theta_2=\beta_M$. The quantities 
$X_j$, $X_j^\prime$, $Y_j$, $Y_j^\prime$, $W_j$ and $W_j^\prime$, 
with $j$ = 
$1$ and $2$, are given by
%with different expressions of $X,~X',~Y,~Y',~W,~W'$ for different phases.\\
%\underline{For $\alpha_M$}
\begin{eqnarray}
X_1&=&A_i-\{D_r\sin(\beta_M-\alpha_M)+D_i\cos(\beta_M-\alpha_M)+F_r\sin2(\beta_M-\alpha_M)+F_i\cos2(\beta_M-\alpha_M)+E_i\},\nonumber\\
X'_1&=&\{D_r\cos(\beta_M-\alpha_M)-D_i\sin(\beta_M-\alpha_M)+F_r\cos2(\beta_M-\alpha_M)-F_i\sin2(\beta_M-\alpha_M)+E_r\}-A_r,\nonumber\\
Y_1&=&A_r+\{D_r\cos(\beta_M-\alpha_M)-D_i\sin(\beta_M-\alpha_M)+F_r\cos2(\beta_M-\alpha_M)-F_i\sin2(\beta_M-\alpha_M)+E_r\},\nonumber\\
Y'_1&=&A_i+\{D_r\sin(\beta_M-\alpha_M)+D_i\cos(\beta_M-\alpha_M)+F_r\sin2(\beta_M-\alpha_M)+F_i\cos2(\beta_M-\alpha_M)+E_i\},\nonumber\\
W_1&=&B_r+C_r\cos(\beta_M-\alpha_M)-C_i\sin(\beta_M-\alpha_M),\nonumber\\
W'_1&=&B_i+C_r\sin(\beta_M-\alpha_M)+C_i\cos(\beta_M-\alpha_M),
\end{eqnarray}
%\underline{For $\beta_M$}
\begin{eqnarray}
X_2&=&A_i-\{D_i\cos(\beta_M-\alpha_M)-D_r\sin(\beta_M-\alpha_M)+E_i\cos2(\beta_M-\alpha_M)-E_r\sin2(\beta_M-\alpha_M)+F_i\},\nonumber\\
X'_2&=&\{D_r\cos(\beta_M-\alpha_M)+D_i\sin(\beta_M-\alpha_M)+E_r\cos2(\beta_M-\alpha_M)+E_i\sin2(\beta_M-\alpha_M)+F_r\}-A_r,\nonumber\\
Y_2&=&A_r+\{D_r\cos(\beta_M-\alpha_M)+D_i\sin(\beta_M-\alpha_M)+E_r\cos2(\beta_M-\alpha_M)+E_i\sin2(\beta_M-\alpha_M)+F_r\},\nonumber\\
Y'_2&=&A_i+\{D_i\cos(\beta_M-\alpha_M)-D_r\sin(\beta_M-\alpha_M)+E_i\cos2(\beta_M-\alpha_M)-E_r\sin2(\beta_M-\alpha_M)+F_i\},\nonumber\\
W_2&=&C_r+B_r\cos(\beta_M-\alpha_M)+B_i\sin(\beta_M-\alpha_M),\nonumber\\
W'_2&=&C_i+B_i\cos(\beta_M-\alpha_M)-B_r\sin(\beta_M-\alpha_M).
\end{eqnarray}
In the above, the suffixes $i$ and $r$ stand for 
imaginary and real part respectively and 
\begin{eqnarray}
 \beta_M-\alpha_M=\cos^{-1}\left[\frac{|(M^{\epsilon_1,\epsilon_2,\delta_{CP}}_\nu)_{11}|^2-c_{12}^4m_1^2-s_{12}^4m_2^2}{2c_{12}^2s_{12}^2m_1m_2}\right].
\end{eqnarray}
Moreover, the complex quantities $A,~B,~C,~D,~E$ and $F$ are defined 
as 
\begin{eqnarray}
A&=&m_3^2[Z-1],\nonumber\\
B&=&m_3m_1\left[Zs_{12}^2(1+t_{23}^4){t_{23}^{-2}}+
Z\sin2\theta_{12}s_{13}e^{i\delta_{CP}}(1-t_{23}^2){t_{23}^{-1}}
+2s_{12}^2\right],\nonumber\\
C&=&m_3m_2\left[Zc_{12}^2(1+t_{23}^4){t_{23}^{-2}}+
Z\sin2\theta_{12}s_{13}e^{i\delta_{CP}}(t_{23}^2-1){t_{23}^{-1}}+
2c_{12}^2\right],\nonumber\\
D&=&m_1m_2\left[2Zc_{12}^2s_{12}^2+Z\sin2\theta_{12}\cos2\theta_{12}s_{13}
e^{i\delta_{CP}}(t_{23}^2-1){t_{23}^{-1}}-2s_{12}^2c_{12}^2\right],\nonumber\\
E&=&m_1^2\left[Zs_{12}^4+Zs_{12}^2\sin2\theta_{12}s_{13}e^{i\delta_{CP}}
(t_{23}^2-1){t_{23}^{-1}}-s_{12}^4\right],\nonumber\\
F&=&m_2^2\left[Zc_{12}^4-Zc_{12}^2\sin2\theta_{12}s_{13}e^{i\delta_{CP}}
(t_{23}^2-1){t_{23}^{-1}}-c_{12}^4\right],
\end{eqnarray}
where $t_{23}\equiv\tan\theta_{23}$ and 
$Z={[(M^{\epsilon_1,\epsilon_2,\delta_{CP}}_\nu)_{23}]^2}
{[{(M^{\epsilon_1,\epsilon_2,\delta_{CP}}_\nu)_{22}
(M^{\epsilon_1,\epsilon_2,\delta_{CP}}_\nu)_{33}}]}^{-1}$.
\section{Numerical Analysis}
 %%%%%%%%%%%   Table I %%%%%%%%%%%%%
\begin{table}[!htb]
\begin{center}
\caption{Input experimental values \cite{GonzalezGarcia:2012sz}}
\begin{tabular}{|c|c|}
\hline
{Quantity} & {$3\sigma$ ranges}\\
\hline
$\Delta_{21}^2$ & $7.00<\Delta_{21}^2(10^{5}~ eV^{-2})<8.09$\\
$\Delta_{32}^2<0$ & $-2.649<\Delta_{32}^2(10^{3}~ eV^{-2})<-2.242$\\
%$\Delta_{32}^2>0$ & $2.276<\Delta_{32}^2(10^{3}~ eV^{-2})<2.695$\\
$\Delta_{32}^2>0$ & $2.195<\Delta_{32}^2(10^{3}~ eV^{-2})<2.625$\\
$\theta_{12}$ & $31.09^\circ<\theta_{12}<35.89^\circ$\\
$\theta_{23}$ & $35.80^\circ<\theta_{23}<54.80^\circ$\\
$\theta_{13}$ & $7.19^\circ<\theta_{13}<9.96^\circ$\\
$\delta_D$ & Unconstrained\\
\hline
\end{tabular}
\label{input}
\end{center}
\end{table}
%%%%%%%%%%%%%%%%%%%%%%%%%%%%%%%%%%%%%%%%%%%%%%%%%%%%%%%%%%%%%%
We analyse four $M_\nu^{\epsilon_1,\epsilon_2,\delta}$'s  from A1, A2, A3, A4, 
and two $M_\nu^{\epsilon_1,\epsilon_2,\delta}$'s from the 
two pairs (B1,B2) and (B3,B4) to determine the 
admitted 
parameter space accommodating all experimental data from 
neutrino oscillation studies. 
We utilise the $3\sigma$ ranges given in Table 5 as inputs to 
constrain the parameter space. Each of the above six categories of textures 
contains ten 
parameters. First of all, {\it the texture A1 is ruled out due to its $\theta_{13}$ 
value being outside the allowed interval}. Except A1, each of the 
 other five $M_\nu^{\epsilon_1,\epsilon_2,\delta}$'s
admits a constrained parameter space.   
We follow a minimalistic approach in which we keep as many  as 
parameters equal to zero as possible. The allowed ranges of the 
parameters are given in Table 6. 
\vskip 0.1in
\noindent
For each category, we present our predictions in Table 7 on the 
individual masses of the three neutrinos and their sum, 
$|m_{\nu_{\beta\beta}}|$ relevant to $0\nu2\beta$ decay, the 
CP violating parameters 
$J_{CP}$ and $\delta_{CP}$ and the Majorana phases $\alpha_M$ and $\beta_M$.
To this end, we note that the testability of each texture crucially depends
 on the 
nature of 
the light neutrino mass ordering which will hopefully 
be experimentally determined in the near future. 
Again, a composite analysis is performed, including 
cosmological and astrophysical experimental data, such as those from the 
recent  
PLANCK satellite, WMAP low-$l$ polarization, 
gravitational lensing and the Hubble constant $H_0$ from 
Hubble space telescope data with priors. 
These imply  a value of $\Sigma m_i<1.11$eV 
whereas the incorporation of the SDSS DR8 results with the above combination 
reduces 
this upper limit drastically to 0.23 eV. 
Altogether, we consider a 
conservative range 
of the upper limit on $\Sigma m_i$ as $\Sigma m_i<(0.23-1.11)$eV 
\cite{Ade:2013zuv,Giusarma:2013pmn}. In our analysis, the 
predicted values of $\Sigma m_i$ in all the cases are far below  
the lower limit of the 
aforesaid range. Another prediction, $|m_{\nu_{\beta\beta}}|$, to be measured 
in neutrinoless double beta decay experiments, 
is also much less than the 
quoted limit $|m_{\nu_{\beta\beta}}|<( 0.14- 0.38)$eV 
\cite{Rodejohann:2012xd,Bahcall:2004ip,Cremonesi:2012av} 
presented by 
the EXO-200 collaboration \cite{Auger:2012ar}. 
Finally, we have also included our CP violating Jarlskog  parameter
$J_{CP}$ and the Dirac CP phase $\delta_{CP}$ \cite{Abe:2011sj} in Table 7.
The former can be extracted from experiments looking for 
CP violation with neutrino and antineutrino beams by
measuring the difference in oscillation probabilities
${\rm P}(\nu_\mu\rightarrow \nu_e)$ --
${\rm P}(\bar{\nu_\mu}\rightarrow {\bar{\nu_e}})$. A detailed review of this 
is
presented in Ref. \cite{Minakata:2008yz}.
An estimation of the latter on the basis of global analysis 
is given in Ref. \cite{GonzalezGarcia:2012sz}
with the result  
$\delta_{CP} = {\left(300^{+66}_{-138}\right)}^o$ for 1$\sigma$ range.
\section{Concluding Summary}
We have considered maximal zero textures in the context of the inverse seesaw 
mechanism. Lepton number violation is posited here through a keV scale 
Majorana mass matrix $\mu$. The observed nontrivial mixing of every light neutrino flavor 
has been used. We have also assumed that none of the light neutrinos is 
massless. Consequently our allowed choices narrowed down to 
eight 4-zero$\times$6-zero, eight 6-zero$\times$4-zero and eight 
4-zero$\times$4-zero combinations in $F=M_D M^{-1}$ which controls neutrino 
masses and mixing angles through the relation $M_\nu = F\mu {F}^T$. The first 
sixteen texture combinations give rise to a null lepton asymmetry of the 
Universe and hence are excluded on that count. The remaining eight texture combinations can effect
 leptogenesis and are tested by all available and relevant 
neutrino oscillation data. Only seven combinations leading to 
five distinct neutrino matrix forms are seen to be allowed with 
strongly constrained parameter spaces. Two of these are seen to imply a 
normal mass ordering of the light neutrinos while the remaining three lead to 
an inverted one. 
\vskip 0.1in
\noindent
{\bf Acknowledgement}
\vskip 0.1in
\noindent
We thank Mainak Chakraborty for computational assistance.  
%%%%%%%%%%%   Table I %%%%%%%%%%%%%
%\begin{table}[!ht]
%\caption{Input experimental values \cite{GonzalezGarcia:2012sz}}
%\begin{tabular}{|c|c|}
%\hline
%{Quantity} & {$3\sigma$ ranges}\\
%\hline
%$\Delta_{21}^2$ & $7.00<\Delta_{21}^2(10^{5}~ eV^{-2})<8.09$\\
%$\Delta_{32}^2<0$ & $-2.649<\Delta_{32}^2(10^{3}~ eV^{-2})<-2.242$\\
%$\Delta_{32}^2>0$ & $2.276<\Delta_{32}^2(10^{3}~ eV^{-2})<2.695$\\
%$\Delta_{32}^2>0$ & $2.195<\Delta_{32}^2(10^{3}~ eV^{-2})<2.625$\\
%$\theta_{12}$ & $31.09^\circ<\theta_{12}<35.89^\circ$\\
%$\theta_{23}$ & $35.80^\circ<\theta_{23}<54.80^\circ$\\
%$\theta_{13}$ & $7.19^\circ<\theta_{13}<9.96^\circ$\\
%$\delta_D$ & Unconstrained\\
%\hline
%\end{tabular}
%\label{input}
%\end{table}
%%%%%%%%%%%%%%%%%%%%%% Table II %%%%%%%%%%%%%%%%%%%%%%%%%
%\narrowtext
\begin{table}
\begin{center}
\begin{tabular}{|c|c|c|c|c|c|}
\hline
Parameters&\multicolumn{5}{|c|}{Categories}\\
\cline{2-6}&
A2&A3&A4&(B1,B2)&(B3,B4)\\
\hline
$k_1$&2.20-2.90&2.80-3.40&1.81-2.00&0.4-4.90&1.50-2.85\\
\hline
$k_2$&2.00-2.95&2.00-2.40&1.60-2.24&0.6-4.90&1.40-2.25\\
\hline
$k_3$&0.55-1.00&0.005-0.010&0.84-2.40&0.21-0.80&0.35-0.55\\
\hline
$\alpha$&89-91&0-180&30-150&10-180&90-180\\
(degree)&&&&&\\
\hline
$\beta$&0&0-160&30-40&10-180&90-160\\
(degree)&&&&&\\
\hline
$\phi_1$&30-100&20-140&90-180&0-180&0\\
(degree)&&&&&\\
\hline
$\phi_2$&0-140&0&0-30&0-180&0\\
(degree)&&&&&\\
\hline
$\epsilon_1$&0.06-0.12&0.06-0.15&0.12-0.15&0.0001-0.15&0.14-0.15\\
\hline
$\epsilon_2$&0.03-0.11&0&0.03-0.12&0.0002-0.15&0.12-0.15\\
\hline
$\delta$&0&0&0&0&0.03-0.09\\
\hline
\end{tabular}
\caption{
Allowed Parameter ranges
}
\end{center}
\end{table} 
%%%%%%%%%%%%%%%%%%%%%% Table III %%%%%%%%%%%%%%%%%%%%%%%%%
%\narrowtext
\begin{table}
\begin{center}
\begin{tabular}{|c|c|c|c|c|c|}
\hline
{\rm Observables}&\multicolumn{5}{|c|}{\rm{Categories}}\\
\cline{2-6}&
A2&A3&A4&(B1,B2)&(B3,B4)\\
\hline
{\rm Hierarchy}&I&I&I&N&N\\
\hline
$m_1$&0.050-0.060&0.049-0.055&0.053-0.061&0.003-0.068&0.017-0.026\\
(eV)&&&&&\\
\hline
$m_2$&0.051-0.061&0.050-0.056&0.054-0.062&0.009-0.069&0.019-0.028\\
(eV)&&&&&\\
\hline
$m_3$&0.016-0.023&0.017-0.021&0.026-0.035&0.048-0.086&0.051-0.058\\
(eV)&&&&&\\
\hline
$\Sigma m_i$&0.115-0.134&0.116-0.132&0.134-0.158&0.061-0.224&0.09-0.113\\
(eV)&&&&&\\
\hline
${m_\nu}_{\beta\beta}$&0.015-0.024&0.016-0.022&0.026-0.035&0.001-0.049&0.005-0.015\\
(eV)&&&&&\\
\hline
$J_{CP}$&0.026-0.036&0.002-0.035&0.002-0.03&3.7$\times {10}^{-5}$-0.0311&0.001-0.03\\
\hline
$\delta_{CP}$&55-90&3-85&3-67&0.1-85&21-90\\
(degree)&&&&&\\
\hline
$\alpha_M$&-57-58&-13-9&-90-90&-90-90&-84-34\\
(degree)&&&&&\\
\hline
$\beta_M$&-90-90&-45-40&-90-90&-90-90&-63-10\\
(degree)&&&&&\\
\hline
\end{tabular}
\caption{
{\rm Predictions}
}
\end{center}
\end{table} 
\appendix
\section{Appendix}
\vskip 0.1in
\noindent
{\bf{Relevant expressions}}
\begin{eqnarray}
&&\theta=\frac{1}{3}\tan^{-1}\left[\frac{3\sqrt{3}a\sqrt{\Delta}}{2b^3-9abc+27a^2d}\right],\nonumber\\
&&r=\sqrt{\left(3\sqrt{3}a\sqrt{\Delta}\right)^2+\left(2b^3-9abc+27a^2d\right)^2},
\end{eqnarray}
where 
\begin{eqnarray}
a&=&1,\nonumber\\
b&=&-(h_{11}+h_{22}+h_{33}),\nonumber\\
c&=&h_{33}h_{11}+h_{33}h_{22}+h_{11}h_{22}-|h_{12}|^2-|h_{13}|^2-|h_{23}|^2,\nonumber\\
d&=&h_{11}|h_{23}|^2+h_{33}|h_{12}|^2+h_{22}|h_{13}|^2-h_{11}h_{22}h_{33}-2 Re(h_{12}h_{23}h_{13}^\ast),
\end{eqnarray}
and 
\begin{equation}
\Delta=18abcd-4b^3d+ b^2 c^2 -4 a c^3 -27 a^2 d^2.
\end{equation}
%%%%%%%%%%%%%%%%%%%%%%%%%%%%%%%%%%%%%%%%%%%


\begin{thebibliography}{99}
\bibitem{Gell}
 M.~Gell-Mann, P.~Ramond and R.~Slansky  in 
'Supergravity', ({\it eds. D.~Friedman and P.~van Nieuwenhuizen}) 
North-Holland, Amsterdam, 315 (1979).

\bibitem{Yana}
T.~Yanagida in Proc. Workshop 
``{\it Unified theory and baryon number in the universe }'', 
(eds. O.~Sawada and A.~Sugamoto KEK, Tsukuba, Japan), 95 (1979).

%\cite{Mohapatra:1979ia}
\bibitem{Mo} 
R.~N.~Mohapatra and G.~Senjanovic,
%  ``{\it Neutrino mass and spontaneous parity violation},''
  Phys.\ Rev.\ Lett.\  {\bf 44}, 912 (1980).
  %%CITATION = PRLTA,44,912;%%

\bibitem{four} 
%\cite{Davidson:2002qv}
%\bibitem{Davidson:2002qv} 
S.~Davidson and A.~Ibarra,
%``A Lower bound on the right-handed neutrino mass from leptogenesis,''
Phys.\ Lett.\ B {\bf 535}, 25 (2002) [hep-ph/0202239].
  %%CITATION = HEP-PH/0202239;%%
  %428 citations counted in INSPIRE as of 14 Nov 2013

\bibitem{Mohapatra:1986aw} 
  R.~N.~Mohapatra,
  %``Mechanism for Understanding Small Neutrino Mass in Superstring Theories,''
  Phys.\ Rev.\ Lett.\  {\bf 56}, 561 (1986).
  %%CITATION = PRLTA,56,561;%%
  %198 citations counted in INSPIRE as of 31 Oct 2013
%\cite{Mohapatra:1986bd}
\bibitem{Mohapatra:1986bd} 
  R.~N.~Mohapatra and J.~W.~F.~Valle,
  %``Neutrino Mass and Baryon Number Nonconservation in Superstring Models,''
  Phys.\ Rev.\ D {\bf 34}, 1642 (1986).
  %%CITATION = PHRVA,D34,1642;%%
  %525 citations counted in INSPIRE as of 31 Oct 2013
%\cite{Bernabeu:1987gr}
\bibitem{Bernabeu:1987gr} 
  J.~Bernabeu, A.~Santamaria, J.~Vidal, A.~Mendez and J.~W.~F.~Valle,
  %``Lepton Flavor Nonconservation at High-Energies in a Superstring Inspired Standard Model,''
  Phys.\ Lett.\ B {\bf 187}, 303 (1987).
  %%CITATION = PHLTA,B187,303;%%
  %219 citations counted in INSPIRE as of 31 Oct 2013

\bibitem{six} 
%\cite{GonzalezGarcia:2009qd}
%\bibitem{GonzalezGarcia:2009qd} 
M.~C.~Gonzalez-Garcia, J.~Racker and N.~Rius,
%``Leptogenesis without violation of B-L,''
JHEP {\bf 0911}, 079 (2009) [arXiv:0909.3518 [hep-ph]].
  %%CITATION = ARXIV:0909.3518;%%
  %18 citations counted in INSPIRE as of 14 Nov 2013

%\cite{Adhikary:2013bma}
%\bibitem{Adhikary:2013bma} 
%  B.~Adhikary, M.~Chakraborty and A.~Ghosal,
%``Masses, mixing angles and phases of general Majorana neutrino mass matrix,''
%  JHEP {\bf 1310}, 043 (2013)
%  [arXiv:1307.0988 [hep-ph]].
%  %%CITATION = ARXIV:1307.0988;%%
%  %1 citations counted in INSPIRE as of 29 Oct 2013
\bibitem{seven}
G. C. Branco, D. Emmanuel-Costa, M. N. Rebelo and P. Roy
%  ``{\it Four zero neutrino Yukawa textures in the minimal seesaw framework},''
{\it Phys.\ Rev.\ D\/} {\bf 77}, 053011 (2008).
 % [arXiv:0712.0774 [hep-ph]].

\bibitem{brancobook}G.~C.~Branco, L.~Lavoura, J.~P.~Silva, 
'{\it{CP\,\, Violation}}'
[International Series of Monographs on Physics], Oxford University Press, 1999.

%\cite{Adhikary:2009kz}
\bibitem{Adhikary:2009kz} 
  B.~Adhikary, A.~Ghosal and P.~Roy,
  %``mu tau symmetry, tribimaximal mixing and four zero neutrino Yukawa textures,''
  JHEP {\bf 0910}, 040 (2009)
  [arXiv:0908.2686 [hep-ph]].
  %%CITATION = ARXIV:0908.2686;%%
  %21 citations counted in INSPIRE as of 31 Oct 2013

%\cite{Adhikary:2010fa}
\bibitem{Adhikary:2010fa} 
  B.~Adhikary, A.~Ghosal and P.~Roy,
  %``Baryon asymmetry from leptogenesis with four zero neutrino Yukawa textures,''
  JCAP {\bf 1101}, 025 (2011)
  [arXiv:1009.2635 [hep-ph]].
  %%CITATION = ARXIV:1009.2635;%%
  %6 citations counted in INSPIRE as of 31 Oct 2013

%\cite{Adhikary:2011pv}
\bibitem{Adhikary:2011pv} 
  B.~Adhikary, A.~Ghosal and P.~Roy,
  %``Neutrino Masses, Cosmological Bound and Four Zero Yukawa Textures,''
  Mod.\ Phys.\ Lett.\ A {\bf 26}, 2427 (2011)
  [arXiv:1103.0665 [hep-ph]].
  %%CITATION = ARXIV:1103.0665;%%
  %3 citations counted in INSPIRE as of 31 Oct 2013

%\cite{Adhikary:2012mt}
\bibitem{Adhikary:2012mt} 
  B.~Adhikary, A.~Ghosal and P.~Roy,
  %``$\theta_{13}$, $\mu\tau$ symmetry breaking and neutrino Yukawa textures,''
  Int.\ J.\ Mod.\ Phys.\ A {\bf 28}, 1350118 (2013)
  [arXiv:1210.5328 [hep-ph]].
  %%CITATION = ARXIV:1210.5328;%%
  %4 citations counted in INSPIRE as of 31 Oct 2013
%\cite{Adhikary:2012zx}
\bibitem{Adhikary:2012zx} 
  B.~Adhikary and P.~Roy,
  %``Neutrino Yukawa textures within type-I see-saw,''
  Adv.\ High Energy Phys.\  {\bf 2013}, 324756 (2013)
  [arXiv:1211.0371 [hep-ph]].
  %%CITATION = ARXIV:1211.0371;%%
  %2 citations counted in INSPIRE as of 31 Oct 2013

\bibitem{mutaureview}
%\cite{Grimus:2012hu}
%\bibitem{Grimus:2012hu} 
W.~Grimus and L.~Lavoura,
%``mu-tau Interchange symmetry and lepton mixing,''
Fortsch.\ Phys.\  {\bf 61}, 535 (2013) [arXiv:1207.1678 [hep-ph]].
%%CITATION = ARXIV:1207.1678;%%
%9 citations counted in INSPIRE as of 14 Nov 2013

%\cite{Adhikary:2013bma}
\bibitem{Adhikary:2013bma} 
  B.~Adhikary, M.~Chakraborty and A.~Ghosal,
%``Masses, mixing angles and phases of general Majorana neutrino mass matrix,''
  JHEP {\bf 1310}, 043 (2013)
  [arXiv:1307.0988 [hep-ph]].
%  %%CITATION = ARXIV:1307.0988;%%
%  %1 citations counted in INSPIRE as of 29 Oct 2013

%\cite{GonzalezGarcia:2012sz}
\bibitem{GonzalezGarcia:2012sz}
M.~C.~Gonzalez-Garcia, M.~Maltoni, J.~Salvado and T.~Schwetz
%''{\it Global fit to three neutrino mixing: critical look at present precision},''
{\it JHEP} {\bf 1212}, 123 (2012).
%\cite{Ade:2013zuv}
\bibitem{Ade:2013zuv} 
  P.~A.~R.~Ade {\it et al.}  [Planck Collaboration],
  %``Planck 2013 results. XVI. Cosmological parameters,''
  arXiv:1303.5076 [astro-ph.CO].
  %%CITATION = ARXIV:1303.5076;%%
  %794 citations counted in INSPIRE as of 29 Oct 2013
%\cite{Giusarma:2013pmn}
\bibitem{Giusarma:2013pmn} 
  E.~Giusarma, R.~de Putter, S.~Ho and O.~Mena,
  %``Constraints on neutrino masses from Planck and Galaxy Clustering data,''
  Phys.\ Rev.\ D {\bf 88}, 063515 (2013)
  [arXiv:1306.5544 [astro-ph.CO]].
  %%CITATION = ARXIV:1306.5544;%%
  %7 citations counted in INSPIRE as of 29 Oct 2013
\bibitem{Rodejohann:2012xd} 
  W.~Rodejohann,
  %``Neutrinoless double beta decay and neutrino physics,''
  J.\ Phys.\ G {\bf 39}, 124008 (2012)
  [arXiv:1206.2560 [hep-ph]].
  %%CITATION = ARXIV:1206.2560;%%
  %36 citations counted in INSPIRE as of 29 Oct 2013
 %\cite{Bahcall:2004ip}
\bibitem{Bahcall:2004ip}
J. N. Bahcall, H. Murayama and C. Pena-Garay
%``{\it What can we learn from neutrinoless double beta decay experiments?}''
  {\it Phys.\ Rev.\ D\/} {\bf 70}, 033012 (2004).
%%CITATION = HEP-PH/0403167;%%
%49 citations counted in INSPIRE as of 29 May 2013
%\cite{Cremonesi:2012av}

\bibitem{Cremonesi:2012av}
O. Cremonesi~O
%``{\it Experimental searches of neutrinoless double beta decay}''
  arXiv:1212.4885 [nucl-ex].
%%CITATION = ARXIV:1212.4885;%%
%1 citations counted in INSPIRE as of 29 May 2013

\bibitem{Auger:2012ar} 
  M.~Auger {\it et al.}  [EXO Collaboration],
  %``Search for Neutrinoless Double-Beta Decay in $^{136}$Xe with EXO-200,''
  Phys.\ Rev.\ Lett.\  {\bf 109}, 032505 (2012)
  [arXiv:1205.5608 [hep-ex]].
  %%CITATION = ARXIV:1205.5608;%%
  %102 citations counted in INSPIRE as of 29 Oct 2013

%\cite{Abe:2011sj}
\bibitem{Abe:2011sj}
  T2K Collab.  (K. Abe\,{\it et al}.), arXiv: 1308.0465 [hep-ex];
%  ``{\it Indication of electron neutrino appearance from an accelerator-produce%d off-axis muon neutrino beam},''
  {\it Phys. Rev. Lett.\/} {\bf 107}, 041801 (2011).

\bibitem{Minakata:2008yz}
H. Minakata
%``{\it Looking for Leptonic CP Violation with Neutrinos},''
{\it Acta Phys.\ Polon.\ B\/} {\bf 39}, 283 (2008).
%\cite{Minakata:2008yz}. 
\end{thebibliography}
\end{document}